# EXPERIMENT FOR TRANSIENT EFFECTS OF SUDDEN CATASTROPHIC LOSS OF VACUUM ON A SCALED SUPERCONDUCTING RADIO FREQUENCY CRYOMODULE

A. Dalesandro[1], J. Theilacker[1], and S. W. Van Sciver[2]

[1] Fermi National Accelerator Laboratory
Batavia, IL, 60510, USA

[2] National High Magnetic Field Laboratory
Florida State University, Tallahassee, FL, 32310, USA

## ABSTRACT

Safe operation of superconducting radio frequency (SRF) cavities require design consideration of a sudden catastrophic loss of vacuum (SCLV) adjacent with liquid helium (LHe) vessels and subsequent dangers. An experiment is discussed to test the longitudinal effects of SCLV along the beam line of a string of scaled SRF cavities. Each scaled cavity includes one segment of beam tube within a LHe vessel containing 2 K saturated LHe, and a riser pipe connecting the LHe vessel to a common gas header. At the beam tube inlet is a fast acting solenoid valve to simulate SCLV and a high/low range orifice plate flow-meter to measure air influx to the cavity. The gas header exit also has an orifice plate flow-meter to measure helium venting the system at the relief pressure of 0.4 MPa. Each cavity is instrumented with Validyne pressure transducers and Cernox thermometers. The purpose of this experiment is to quantify the time required to spoil the beam vacuum and the effects of transient heat and mass transfer on the helium system. Heat transfer data is expected to reveal a longitudinal effect due to the geometry of the experiment. Details of the experimental design criteria and objectives are presented.

## INTRODUCTION

Project X is a proposed linear accelerator to be built at Fermi National Accelerator Laboratory. The accelerator uses SRF cryomodules (strings of 8-10 cavities) to accelerate particles to nearly the speed of light. To maintain superconductivity, the cavities are engulfed in baths of superfluid helium (He II) between 1.8 and 2.0 K. For cryogenic systems of this magnitude, SCLV is largely considered the effective worst-case failure scenario. Catastrophic leak from atmosphere to a cryogenically-cooled vacuum space allows room temperature air to condense on the cold surface, transferring large amounts of heat to the surface and subsequently to the LHe. This heat flux acts to rapidly vaporize the helium and pressurize the cavity. Accommodation of this scenario results in a very large and complex helium relief system.

An important characteristic of this process is the heat flux from room temperature air to the LHe-cooled surface, since this value affects the helium boil-off rate and thus the pressurization of the cavity. Previous work exists on the subject, but experimental geometries tend to be subjective, and are generally concerned with the specific relief requirements of the system being tested. From previous studies of SCLV on LHe systems, a conservative peak heat flux of 4.0 W/cm$^2$ has been established [2-4]. The most relevant study to the current experiment is a test of 1.3 GHz cavities from the German Electron-Synchrotron (DESY) where a full-scale functioning cryomodule was bench tested to determine if the relief system had been sized adequately [1]. The result of that test was positive for DESY but from its data arose some interesting results. In a venting test of beam tube vacuum, it took almost 4 seconds for the pressure in the entire beam length to begin increasing and nearly 18 seconds for the entire tube pressure to equalize with atmosphere. Also the published heat flux from this test is 2.3 W/cm$^2$, almost a factor of two less than the accepted value, and much more favorable for designers of large-scale liquid cryogenic systems that are particularly long relative to flow cross section. Unfortunately the assumed error in this heat flux is ± 50%, which on the high side is 3.45 W/cm$^2$, essentially not much better than the accepted value. However given the surprising pressure lag in the data, it is possible that the effective heat load to the cryogenic system might be significantly reduced by this longitudinal effect. A designated worst-case heat flux of this magnitude could significantly reduce the cost and complexity of safely relieving the pressure for a large system, such as Project X. It is the primary goal of the experiment described herein to measure this longitudinal effect and how it acts to reduce the effective heat load to the cryogenic system.

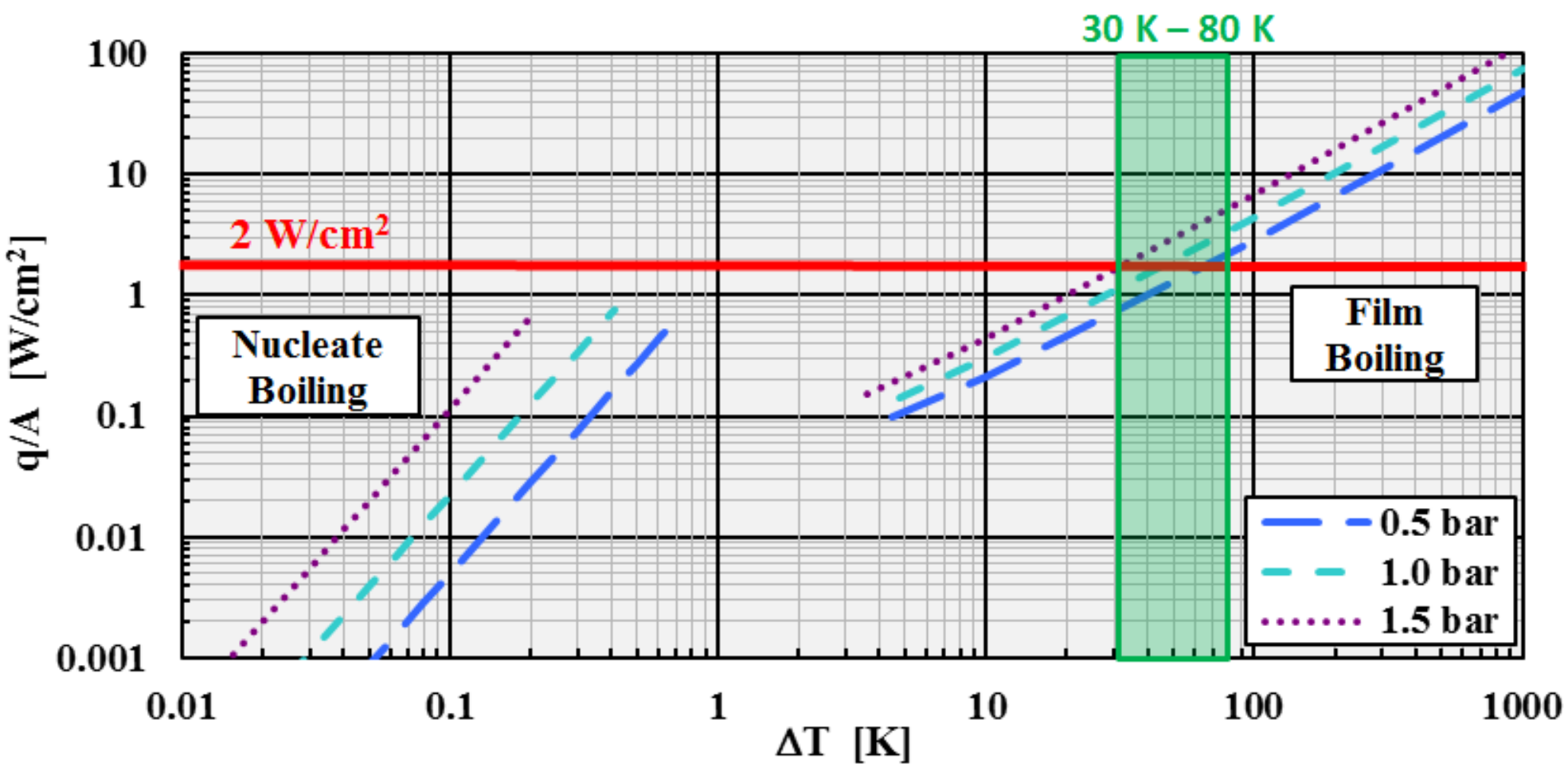

**FIGURE 1.** Modified Plot of Pool Boiling Heat Transfer for He-I [6].

There are additional data to support the claim of a heat flux on the order of 2 W/cm$^2$, shown in FIGURE 1. This is a modified plot of data produced from equations provided in existing literature for a horizontal cylinder with outer diameter greater than 1 cm (applicable to the current experiment) from normal fluid helium (He I) [6]. In the figure, the range of temperature differences ($\Delta$T) from 30-80 K is highlighted, emphasizing its intersection with film boiling heat transfer at a given pressure to a corresponding heat flux, in this case 1-4 W/cm$^2$. Given the triple point and saturation temperatures for nitrogen of 63 K and 77 K respectively, and considering a finite $\Delta$T through the thickness of a beam tube of around 20 K, the $\Delta$T between the bulk LHe and the beam surface will be between 38 K and 55 K. If no $\Delta$T exists through the beam thickness, the maximum $\Delta$T to the LHe is 75 K, and if the $\Delta$T across the thickness is higher, this helps to reduce the film boiling heat flux. This heat flux is consistent with air condensation on LHe cooled surfaces. It is important to note that in He I the peak heat flux is due to film boiling, but in He II there is a strong thermal conductivity mechanism such that this plot is not applicable. A lumped system model, calculates a time delay of 0.5 second for the LHe pressure in the experiment to exceed the lambda transition, although the true time is likely greater than this given the shortcomings of the model, see below.

## DESIGN PROCESS

### Basic Modeling

The first step in this experiment was a proof of concept approach to ensure that the desired data could be measured accurately and reliably. An experiment was built and instrumented to measure temperature and pressure in a single scaled cavity, with the beam tube pumped to vacuum and cooled to 2 K, then rapidly vented to atmosphere. The results showed flaws in choice of instrumentation and some poor mounting techniques, but nevertheless produced repeatable data for fundamental temperatures and pressures as well as verification of relevant timescales [5]. This experiment also provided valuable insight leading to the current experiment.

The current experiment seeks to expand on the data produced in previous work, particularly on interesting SCLV mechanisms [1-5]. The primary objective of the experiment is to measure and characterize the longitudinal effect, as noted by DESY [1]. Additionally the experiment is instrumented to measure temperature data relevant to heat transfer through the beam tube, to characterize the dominant heat transfer regimes versus length throughout the span of the SCLV process, discussed later. The design of the experiment is as accommodating to these criteria as possible, given size constraints of the cryostat, and relevant scaling parameters shown in TABLE 1. Since many interesting SCLV mechanisms are geometry dependent, it is desirable to scale the experiment as closely to a realistic cavity as possible. It may be evident from TABLE 1, which shows some relevant scaling factors for several SCLV experiments, that the single cavity test was scaled closely to a 1.3 GHz cavity, whereas the string experiment deviates from this goal by increasing the liquid to vapor helium ratio, for two reasons. The primary reason is to encourage the system to reach the relief set point of 0.4 MPa such that higher pressure supercritical data is collected, which did not happen in the single cavity experiment due in large part to a leaking o-ring. Another reason is to encourage a larger amount of data per run of the experiment related to heat transfer in the liquid regime.

**TABLE 1.** Scaling Parameters for SCLV Experiment Relative to Previous Experiments [1,2,4,5].

| | | DESY | AMS | CEBAF | 1.3 GHz | Single Cavity | Cavity String |
|---|---|---|---|---|---|---|---|
| $T_{Liq}$ | K | 2 | 1.907 | 2 | 2 | 2 | 2 |
| $V_{Liq}$ | L | 150 | 10.8 | 282.6 | 150 | 0.44 | 7.126 |
| $V_{Vap}$ | L | 850 | 3 | 127.9 | 850 | 2.83 | 21.175 |
| $P_{relief}$ | MPa | 0.196 | 1.150 | 0.120 | 0.39 | 0.40 | 0.40 |
| $A_{surf}$ | $m^2$ | 6.72 | 0.314 | 1.61 | 6.72 | 0.024 | 0.1357 |
| $V_{Liq}/V_{He}$ | - | 0.150 | 0.783 | 0.688 | 0.150 | 0.135 | 0.252 |
| $A_{surf}/V_{He}$ | $cm^2/L$ | 67.20 | 227.5 | 39.22 | 67.20 | 73.39 | 47.95 |
| $A_{surf}/V_{Liq}$ | $cm^2/L$ | 448.0 | 290.7 | 56.97 | 448.0 | 545.5 | 190.5 |

To estimate what kind of results might actually be generated by a geometry dependent system, a good starting point is a lumped system analysis of the experiment, mentioned above. In this type of analysis all of the LHe is lumped together as a common volume, as is the helium gas, and based on the starting conditions of 2.0 K and 3.13 kPa (saturated) the initial mass and energy of the system, as well as other helium state properties are determined by the code HEPAK [7]. By assuming a constant mass and volume system (true until relief) and isothermal conditions (not likely, but very difficult otherwise), the bulk density is 37.3 kg/m$^3$, which refers to the total mass of helium by the total volume of the experiment. By setting a fixed relief pressure at 0.4 MPa and using the bulk density, the approximate relief temperature is 7.4 K. Knowing the relief pressure and temperature provides the total energy of the fluid at the relief condition, and if the heat flux is assumed constant, the time for the system to reach the relief point is calculated using a simple energy balance. At an assumed heat flux of 2 W/cm$^2$ the time to reach the relief point for the string is approximately 11 seconds, shown in TABLE 2. FIGURE 2 compares results from this model to published results from several SCLV experiments using the heat flux published by each respective experiment, to illustrate consistency of the model. The discrepancy between the model results and data for CEBAF 2 K is because when that experiment started to relieve, the helium in the experiment was still two-phase, which significantly increases the difficulty of calculating a bulk density.

**TABLE 2.** Lumped System Pressurization Analysis – String Experiment.

| Initial Conditions | | | | | | | | | | | | | | | | | |
|---|---|---|---|---|---|---|---|---|---|---|---|---|---|---|---|---|---|
| $T_i$ | 2.0 | K | $V_{liq}$ | 7.126 | L | $\rho_{Liq}$ | 145.7 | kg/m$^3$ | $M_{liq}$ | 1.038 | kg | $u_{Liq}$ | 1.62 | J/g | $U_{liq}$ | 1,682.2 | J |
| $P_i$ | 3,129 | Pa | $V_{vap}$ | 21.175 | L | $\rho_{Vap}$ | 0.794 | kg/m$^3$ | $M_{vap}$ | 0.017 | kg | $u_{Vap}$ | 21.10 | J/g | $U_{vap}$ | 354.5 | J |
| | | | $V_{tot}$ | 28.301 | L | $\rho_{bulk}$ | 37.27 | kg/m$^3$ | $M_{tot}$ | 1.055 | kg | $h_{fg}$ | 23.05 | J/g | $U_{Tot,i}$ | 2,036.8 | J |
| | | | **Relief Conditions** | | | | | | | | | | | | | | |
| | | | $T_{relief}$ | 7.36 | K | $u_{relief}$ | 30.52 | J/g | Q/A | 20,000 | W/m$^2$ | | | | | | |
| | | | $P_{relief}$ | 405,300 | Pa | $U_{Tot,f}$ | 32,185 | J | A | 0.1357 | m$^2$ | $t_{relief}$ | 11.1 | s | | | |
| | | | $\rho_{bulk}$ | 37.27 | kg/m$^3$ | $\Delta U_{Tot}$ | 30,149 | J | Q | 2,714 | J/s | | | | | | |

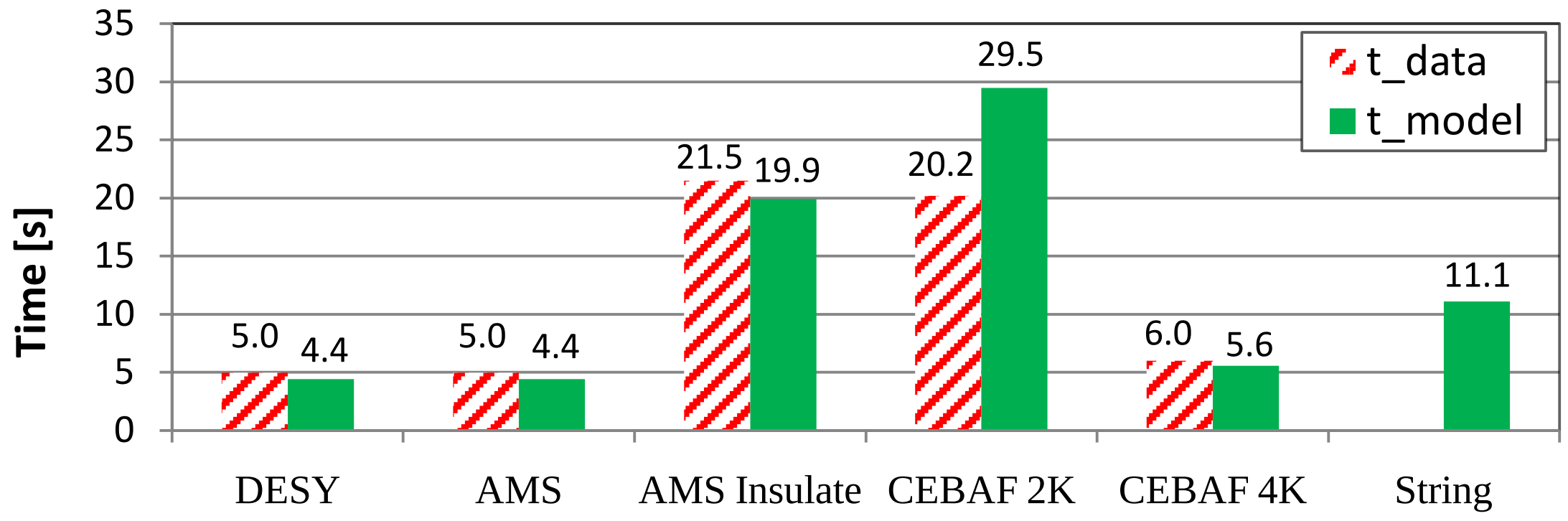

**FIGURE 2.** Time to Reach Relief Pressure, Published Data vs. Model [1,2,4].

This relief scenario is somewhat of an approximation, however, since the above model treats the experiment as a lumped system, including the total surface area of the beam tube as the surface area of a large, homogenous helium vessel. It also assumes a constant, uniform heat flux. In reality the cold beam tube surface area is more discretized by individual sections in contact with each helium vessel, and due to its length vs. diameter a longitudinal effect is likely. This affects the result of the model in two ways: first it makes the assumption of an isothermal system less likely, and secondly it acts to reduce the effective heat load to the helium, thus increasing the amount of time required to reach the relief pressure. This effect is desirable from a relief standpoint, allowing the system to be relieved by a smaller valve.

While the study of the longitudinal effect is paramount in this experiment, it is perhaps just as important to understand exactly its cause, if its effect is to be used to a system designer's advantage. The most obvious explanation is a cryopumping effect, where gas atoms from a recent influx of atmosphere condense onto an extremely cold surface. This is a real effect and can be observed in almost any vacuum system exposed to low temperatures. Another possibility is that the atmosphere freezing on the inner surface of the beam forms either a plug, or a reduced cross-section that restricts flow. While it is difficult to assume that a given cross-section plugs entirely, it is likely that enough freezing occurs to restrict flow. The significance of this effect however is unknown, but worth considering in the results. Placement of pressure transducers along the length of the beam tube allows for evidence if the flow is plugged, constricted significantly, or if this scenario plays a negligible role in the longitudinal effect.

## Design Layout

The layout of the string experiment, shown in FIGURE 3, includes six scaled cavities each with a dedicated LHe chamber. The cavities share a common beam tube which is initially under vacuum internally and plumbed out of the cryostat on the near end with a warm air flow meter external to the cryostat. Each cavity has a LHe chamber connected to a common helium gas header via a riser pipe. The helium gas header is plumbed on the near end through a cold helium flow meter and out of the cryostat. To cool-down, the LHe chambers are filled at atmospheric pressure via individual 4.7 mm (3/16 inch) fill lines connected to a common 6.4 mm (1/4 inch) fill line from a large reservoir at the far end of the string (not pictured). Once the LHe chambers are all full, the reservoir supply valve is closed before the helium is pumped to saturated vapor pressure at 2.0 K (p = 3.1 kPa). During cool-down, data is collected at a slow rate for monitoring (~Hz). Once the liquid level and experimental pressures and temperatures are at steady state, a fast acting 1 inch solenoid valve upstream of the warm air flow meter is opened and data is collected at a fast sample rate (~kHz). The data acquisition (DAQ) for this experiment is done using National Instrument (NI) Signal Express software and a USB-6225 DAQ with 80 differential inputs and a maximum aggregate sample rate of 250 kHz. Pressure signal conditioning is done using a Validyne MC1-10 carrier demodulator which provides a ±10 VDC output at full range. The cryostat has both an 80 K shield and a 4 K shield to help the experiment reach a steady state temperature more easily and to reduce ambient heat leak to the helium.

This experiment collects temperature and pressure (absolute and differential) data only. All temperature sensors are Lakeshore CX1050-SD thermometers which are specifically made for measurements at low temperatures, and are mounted to stainless steel via indium solder for a robust mechanical and thermal bond. All pressure sensors are Validyne DP10 differential pressure transducers which have shown reliability at low temperatures and are welded to 6.4 mm (1/4 inch) stainless steel tubing or mounted to VCR fittings. To measure the longitudinal pressure effect, seven 140 kPa (20 psi) Validyne pressure transducers are placed approximately 35 cm apart along the length of the beam to measure beam tube pressure relative to cryostat vacuum (~$10^{-9}$ mbar) between each cavity segment. Seven Lakeshore CX1050-SD thermometers are placed along the internal surface of the beam tube to provide temperature data relating to the longitudinal effect. One 55 kPa (80 psi) pressure transducer measures the pressure in the helium gas header and six more measure the LHe chamber pressures (one per chamber) relative to cryostat vacuum. Two thermometers are placed in each LHe chamber: one on the bottom inner surface to provide local bath temperatures as well as provide insight as to when the liquid is gone, and the second is mounted on the surface of the beam tube to provide a temperature gradient across the beam tube thickness, using the internal thermometer. A superconducting liquid level probe is used in the riser pipe of the first and last cavities to ensure that initial liquid level is uniform throughout the string. A 20 kPa (3.2 psi) pressure transducer is connected differentially across the first and last LHe chambers to detect any differential across the string during the experiment. A thermometer is mounted on the upstream face of the cold helium flow meter to help determine local fluid density. Differential pressure measurements across the cold helium flow meter use a 20 kPa transducer, and the warm air flow meter uses both a 20 kPa and a 140 kPa transducer to capture high and low range flow rate data.

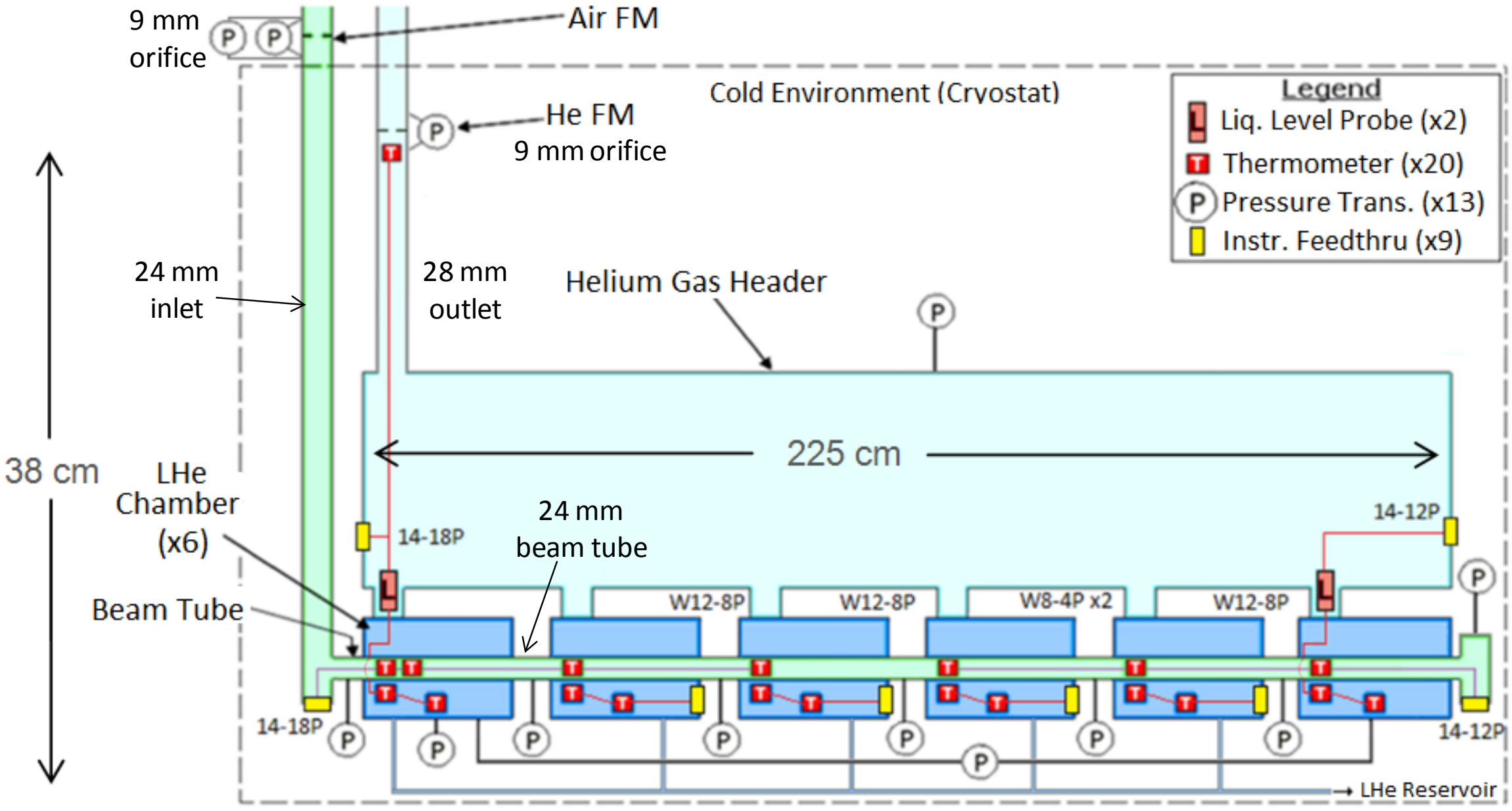

**FIGURE 3.** String Design Schematic.

## Heat Transfer Analysis

A SCLV experiment is essentially a heat transfer problem, where air condensation on a cold surface is vaporizing LHe and pressurizing a helium circuit. The dominate heat transfer mechanism in this process is as much a function of the phase and state of helium as is the temperature gradient. Initially when the LHe is superfluid, the dominant heat transfer mechanism is liquid conduction, until the LHe transitions to normal fluid at which point film boiling dominates heat transfer, and when the helium vaporizes the dominate mechanism becomes free convection in the gas. Generally once a boiling liquid completely vaporizes and natural convection takes over, heat transfer diminishes significantly. However in the case of helium gas at the critical point there is evidence that heat transfer by natural convection is still sufficiently interesting given the state of the helium. In data provided from the SCLV experiment for the Alpha Magnetic Spectrometer, where the pressure and temperature of a helium bath was plotted versus time, the pressurization rate of the helium increased from ~0.1 MPa/s (subcritical) to ~1 MPa/s (supercritical) [2]. This rapid increase in pressurization is the result of an increase in the helium expansivity in the supercritical regime, fueled by heat transfer to the helium due primarily to natural convection.

Designing a thermodynamic experiment to be representative of a future application requires that key heat transfer parameters are comparable. In the case of 1.3 GHz niobium cavities, the wall thickness is 3 mm. Assuming a heat flux of 2 $W/cm^2$, and an average cavity temperature of 50 K, Fourier's law of heat conduction predicts the temperature difference across the thickness to be approximately 0.9 K. To get the same temperature difference across a beam tube made of 304 series stainless steel, the required wall thickness needed is 0.0066 mm. Clearly this is not a practical wall thickness, so to reduce complexity the beam tube has a 0.889 mm (0.035 in) wall thickness (the smallest nominal U.S. standard tube thickness). The thermal resistance for stainless steel at this thickness and the above conditions is greater than 3mm niobium by approximately a factor of three. Subsequently the time required for stainless steel at this thickness to equalize both of its surface temperatures (ignoring axial conduction) is also a factor of three larger than niobium given the above conditions at a given temperature gradient.

To characterize the heat flux and related helium boil-off rate, traditionally it is ideal to quantify the mass of air entering the system (energy in) and the mass of helium venting the system (energy out), as well as the time for the helium bath to reach the relief pressure. A warm air flow meter and a cold helium flow meter, as well as several absolute pressure gauges on the helium circuit can supply this information. If these data are reliable, a simple energy balance can predict the heat flux to the experiment by analyzing the remaining mass and if the temperature and pressure are known. Unfortunately this is a much more complex problem than has just been described for several reasons including temperature differences in the helium due to longitudinal and spatial effects and ambient heat leak, precise mass flow rate measurements through an orifice plate during unsteady flow of a non-ideal gas, and uncertainty in instrumentation.

The method of measuring the heat flux described above is difficult in large part because it is an indirect method to calculate the net energy gain of the system. To help improve the likelihood of measuring an accurate heat flux, this experiment also has the potential to directly measure the $\Delta T$ across the thickness of the beam tube from which a heat flux can be directly calculated using Fourier's law of conduction. The heat transfer through the beam wall thickness is the same as the heat transfer to the helium, ignoring boundary effects and film boiling layers. Since He II has such an anomalously high effective thermal conductivity, these effects can be ignored in this regime. Even in the He I regime where film boiling induces a temperature gradient between the wall and the LHe, which reduces the $\Delta T$ across the wall thickness, the heat flux value calculated can be compared to that calculated by the energy balance method.

## SUMMARY

This experiment is designed to observe and characterize relevant heat transfer data for SRF cavities due to SCLV, specifically related to the longitudinal pressure effect. The longitudinal effect represents a transient pressure front along the length of the beam tube, reducing the rate of air condensation and as such the heat transfer to the LHe. Data of the longitudinal effect will provide a practical understanding of the significance this effect has on SCLV heat transfer in SRF cavities and related geometries. The overall experimental design, including selection and placement of instrumentation, is such to collect data as reliably as possible as per this objective.

## ACKNOWLEDGEMENT

Fermilab is operated by Fermi Research Alliance, LLC under Contract No. DE-AC02-07CH11359 with the United States Department of Energy. Thanks to Mark Vanderlaan for technical assistance.